\documentstyle[preprint,aps,epsf]{revtex}
\begin{document}
%\bibliographystyle {alpha}
%\preprint{WIS--95.--PH}
\draft
\title{
{}From the Zero--bias Anomaly to the Coulomb Blockade:\\
an Exactly Solvable Model.
}
\author{
Alex Kamenev and Yuval Gefen
}
\address{
Department of Condensed Matter, The Weizmann Institute of Science,
Rehovot 76100, Israel
}

\maketitle

\begin{abstract}

A microscopic theory of zero wavelength $({\bf q}=0)$ interaction in
finite--size
systems is proposed. Its exact solution interpolates between the Coulomb
blockade and the perturbative Altshuler--Aronov theory, in the strong and
weak interaction limits respectively. The tunneling density of states and the
quasiparticle life--time are calculated. The physical nature of the
$({\bf q}=0)$ component of the interaction is discussed.

\end{abstract}


e

Much of the activity in the field of mesoscopic physics has focused on
zero--dimensional $(d=0)$ weakly disordered systems, where the relevant energy
scales turn out to be smaller than the Thouless energy, $E_c=\hbar D/L^2$,
$D$ being the diffusivity and $L$ the system's linear size (for a review
see Ref. \cite{AWL}). Studies of
transport as well as of thermodynamic properties reveal the importance  of
electron--electron interactions \cite{Ambegaokar91}.
The role of the latter and the interplay
with disorder have been underlined by the perturbation theory developed by
Altshuler and Aronov (AA) \cite{Altshuler79,Altshuler85}, who consequently
explained the zero--bias anomaly
in tunneling into a {\em macroscopic} conductor. Some ten years ago the
observability of single electron effects in small capacitance devices has
been proposed \cite{Gefen84,Likharev91}, leading to strong suppression of
tunneling at low external
bias. Nazarov \cite{Nazarov}
has pointed out to the common physics between the perturbative
AA anomaly and the non--perturbative Coulomb blockade.

There are several interesting issues pertaining to $d=0$ systems. One is the
tunneling density of states (DOS). To study this quantity it is possible to
employ perturbation theory in the spirit of AA; in doing so one should recall
that the relevant wavenumbers are quantized with a special role played by the
${\bf q}=0$ component of the interaction. This component, pertaining to
fluctuations of the global charge, is unscreened. When the interaction
associated with this component is strong the phenomenological Coulomb blockade
theory may be employed, but it is also desirable to have a microscopic theory
at hand. Another interesting issue involves the electron inelastic
life--time.

Here we present an {\em exactly solvable microscopic} model for interacting
electrons in zero dimensional systems. Our model interpolates between the
Coulomb blockade limit and the zero--bias anomaly. The results for the latter
turn out to be in line with a simple generalization of the AA calculations
\cite{Altshuler85}.
We thus find the tunneling DOS as function of energy, temperature and
coupling strength. We also evaluate the single particle life--time,
associated with the width of the single particle levels. Finally we discuss
the nature of the ${\bf q}=0$ interaction term, $V({\bf q}=0)$. Relating it to
the
system's capacitance we are able to establish a connection with previous
studies of current fluctuations through an ultrasmall capacitance junction
\cite{Devoret91,Glazman91}.

We consider a system described by the Hamiltonian
\begin{equation}
H=\sum_{\alpha}\epsilon_{\alpha}a^{+}_{\alpha}a_{\alpha}+
\frac{V}{2}\left( \sum_{\alpha}a^{+}_{\alpha}a_{\alpha} -N_0 \right)^2,
                                                            \label{H}
\end{equation}
where $\{\epsilon_{\alpha}\}$ represent exact single particle eigenenergies;
the parameter $N_0$ depends on the positive background and the external
fields (e.g. gate voltages), and needs not to be an integer.
The imaginary time single particle Green function is given by \cite{Negele}
\begin{equation}
{\cal G}_{\alpha}(\tau_1,\tau_2,\mu)=-\frac{1}{Z(\mu)}
\int{\cal D}[\psi^{*}_{\alpha}(\tau)\psi_{\alpha}(\tau)]
e^{-\int_0^{\beta}d\tau \left[
\sum_{\alpha}\psi^{*}_{\alpha}(\partial_{\tau}-\mu) \psi_{\alpha}
+H[\psi^{*}_{\alpha},\psi_{\alpha}] \right] }
\psi_{\alpha}(\tau_1)\psi^{*}_{\alpha}(\tau_2),
                                                         \label{G}
\end{equation}
where $Z(\mu)$ is the  partition function, given by the same integral without
pre--exponential  factors.
Next we perform Hubbard--Stratonovich transformation by
means of the imaginary Bose field $\sigma(\tau)$, yielding
\begin{eqnarray}
{\cal G}_{\alpha}(\tau_1,\tau_2,\mu)=-\frac{1}{Z(\mu)}
&&\int{\cal D}[\sigma(\tau)]
e^{\int_0^{\beta}d\tau \left[
\frac{\sigma^2(\tau)}{2V}+\sigma(\tau) N_0 \right]}    \nonumber \\
&&\int{\cal D}[\psi^{*}_{\alpha}\psi_{\alpha}]
e^{-\int_0^{\beta}d\tau
\sum_{\alpha}\psi^{*}_{\alpha}
(\partial_{\tau}-\mu+\epsilon_{\alpha}+\sigma(\tau)) \psi_{\alpha} }
\psi_{\alpha}(\tau_1)\psi^{*}_{\alpha}(\tau_2)     \nonumber \\
=\frac{1}{Z(\mu)}&&\int{\cal D}[\sigma(\tau)]
e^{\int_0^{\beta}d\tau \left[
\frac{\sigma^2(\tau)}{2V}+\sigma(\tau) N_0 \right]}
Z^{[\sigma]}(\mu){\cal G}^{[\sigma]}_{\alpha}(\tau_1,\tau_2,\mu)
                                                          \label{HS}
\end{eqnarray}
with the same transformations in $Z(\mu)$. Here $Z^{[\sigma]}(\mu)$ and
${\cal G}^{[\sigma]}_{\alpha}(\tau_1,\tau_2,\mu)$ are the partition function
and
the Green function, respectively, in the presence of a time
dependent spatially uniform  potential, $\sigma(\tau)$, superimposed on the
single particle problem \cite{Fin}.
Introducing the Matsubara representation (for the
boson field $\omega_m=2\pi i m T,\,\, m=0,\pm 1,\pm 2\ldots $),
\mbox{$\sigma_m=\beta^{-1}\int_0^{\beta}d\tau\sigma(\tau)\exp\{\omega_m\tau\}$
}, we obtain
\begin{eqnarray}
&&Z^{[\sigma]}(\mu)=Z^{[0]}(\mu-\sigma_0)        \nonumber  \\
&&{\cal G}^{[\sigma]}_{\alpha}(\tau_1,\tau_2,\mu)=
{\cal G}^{[0]}_{\alpha}(\tau_1,\tau_2,\mu-\sigma_0)
e^{\int_{\tau_1}^{\tau_2}d\tau [\sigma(\tau)-\sigma_0]},
                                                          \label{ZG}
\end{eqnarray}
where $Z^{[0]}(\mu)$
and ${\cal G}^{[0]}_{\alpha}(\epsilon_n,\mu)$ are partition function and Green
function of non--interacting electrons. Substituting in Eq.(\ref{HS}), we
have
\begin{eqnarray}
{\cal G}_{\alpha}(\tau_1,\tau_2,\mu)=\frac{1}{Z(\mu)}
&&\int d\sigma_0
e^{\beta[\frac{\sigma_0^2}{2V}+\sigma_0 N_0-\Omega^{[0]}(\mu-\sigma_0)]}
{\cal G}^{[0]}_{\alpha}(\tau_1,\tau_2,\mu-\sigma_0)    \nonumber  \\
&&\int\prod_{m\neq 0} d\sigma_m
e^{\sum_{m\neq 0} \left[ \frac{\sigma_m\sigma_{-m}}{2VT}
-\frac{\sigma_m}{\omega_m}(\exp\{-\omega_m\tau_2\}-\exp\{-\omega_m\tau_1\})
\right] }.
                                                          \label{G1}
\end{eqnarray}
Here $\Omega^{[0]}$ is the (non--interacting) thermodynamic potential.
The integral over the static component, $\sigma_0$, describes the smooth
transition between the grand canonical ensemble with the chemical potential
$\mu$ ($V=0$) and the canonical ensemble with a given number of particles
$[N_0]$ ($V=\infty$). For sufficiently  large  systems (or high temperature),
$\frac{1}{\Delta}+\frac{1}{V}\gg\beta$
($\Delta^{-1}\equiv-\partial^2\Omega^{[0]}/\partial\mu^2$ -- is the mean level
spacing), differences between the two ensembles are negligible
and the integral
over $\sigma_0$ may be evaluated within a saddle point approximation
(at $\sigma_0=\overline\sigma_0$).
The $m\neq 0$ components are  readily evaluated. The single particle
Green function assumes the form
\begin{equation}
{\cal G}_{\alpha}(\tau_1-\tau_2,\mu)=
{\cal G}^{[0]}_{\alpha}(\tau_1-\tau_2,\mu-\overline\sigma_0)
{\cal D}(\tau_1-\tau_2),
                                                          \label{GT}
\end{equation}
where the Green function of the  auxiliary field $\sigma$, representing the
{\em Coulomb boson}, is given by
\begin{equation}
{\cal D}(\tau)=
e^{-\frac{V}{2}\left(|\tau|-\frac{\tau^2}{\beta} \right)}; \,\,\,\,
|\tau|\leq\beta
                                                          \label{D}
\end{equation}
(it is periodic with a period $\beta$).
In the Matsubara representation the Green function is superimposed with the
Coulomb boson, see Fig. \ref{fg:1}, to yield
\begin{equation}
{\cal G}_{\alpha}(\epsilon_n,\mu)=T\sum_{\omega_m}
{\cal G}^{[0]}_{\alpha}(\epsilon_n-\omega_m,\overline\mu)
{\cal D}(\omega_m),
                                                          \label{GM}
\end{equation}
where $\overline\mu\equiv\mu-\overline\sigma_0$,
$\epsilon_n=2\pi i(n+1/2)T$ and ${\cal D}(\omega_m)$ is
evaluated as the Fourier transform of ${\cal D}(\tau)$ (in imaginary time),
over the period $[0,\beta]$. The related spectral function is defined by
$B(\omega)\equiv 2\Im{\cal D}^A(\omega)$ \cite{Abrikosov64}
(${\cal D}^A$ is the advanced Green
function of the Coulomb boson) and is given by
\begin{equation}
B(\omega)=\sqrt{\frac{2\pi}{VT}} \left(
e^{-\frac{(\omega+V/2)^2}{2VT} }-
e^{-\frac{(\omega-V/2)^2}{2VT} } \right).
                                                          \label{B1}
\end{equation}
Employing the spectral representation of ${\cal G}^{[0]}_{\alpha}$,
$A^{[0]}_{\alpha}$, the total Green function assumes the following Lehmann
representation
\begin{equation}
{\cal G}_{\alpha}(\epsilon_n,\mu)=-\frac{1}{2}\int\int_{-\infty}^{\infty}
\frac{d\epsilon^\prime}{2\pi}\frac{d\omega^\prime}{2\pi}
B(\omega^\prime)A^{[0]}_{\alpha}(\epsilon^\prime,\overline\mu)
\frac{\mbox{coth}\frac{\omega^\prime}{2T}+
\mbox{tanh}\frac{\epsilon^\prime}{2T} }
{\epsilon_n-\omega^\prime-\epsilon^\prime}.
                                                          \label{GM1}
\end{equation}
We are now in a position to evaluate the DOS,
$\nu(\epsilon)\equiv-\pi^{-1}\sum_{\alpha}\Im{\cal G}^{R}_{\alpha}(\epsilon)$.
Analytically continuing ${\cal G}_{\alpha}$, we finally find
\begin{equation}
\nu(\epsilon)=\frac{1}{2}\int_{-\infty}^{\infty}
\frac{d\omega}{2\pi}
B(\omega)
\left( \mbox{tanh}\frac{\omega-\epsilon}{2T}-
\mbox{coth}\frac{\omega}{2T} \right)
\nu^{[0]}(\epsilon-\omega),
                                                          \label{nu}
\end{equation}
where $\nu^{[0]}$ is the DOS related to
${\cal G}^{[0]}_{\alpha}$.
For non--interacting electrons (with or without disorder),
$\nu^{[0]}(\epsilon)\approx \mbox{const}$, substituting Eq.\ (\ref{B1}) into
Eq.\ (\ref{nu}), we obtain
\begin{equation}
\frac{\nu(\epsilon)}{\nu^{[0]}} =\sqrt{\frac{2\pi}{VT}}
e^{-\frac{V}{8T}} \mbox{cosh}\frac{\epsilon}{2T}
\int_{-\infty}^{\infty}
\frac{d\omega}{2\pi}\frac{e^{-\frac{\omega^2}{2VT}}}
{\mbox{cosh}\frac{\omega-\epsilon}{2T}} \,\, .
                                                          \label{nu1}
\end{equation}
Note that this result is independent of the disorder in the system.
We have thus derived an expression for the DOS which interpolates between the
perturbative zero bias anomaly and the semiclassical Coulomb blockade. In the
limit of weak interaction  ($V\ll T$) this reduces to
\begin{equation}
\frac{\nu(\epsilon)}{\nu^{[0]}} =1-\frac{V}{4T}
\mbox{cosh}^{-2}\frac{\epsilon}{2T}.
                                                          \label{nul}
\end{equation}
This turns out to be the zero--bias anomaly of AA if one substitutes
$d=0$ in their analysis \cite{Altshuler85}.
For strong interaction  ($V\gg T$)
\begin{equation}
\frac{\nu(\epsilon)}{\nu^{[0]}} =\left\{ \begin{array}{ll}
{\displaystyle \sqrt{\frac{2\pi T}{V}} \,
e^{-\frac{V}{8T}} \mbox{cosh}\frac{\epsilon}{2T}  }; \,\,\,\,
&\epsilon\ll \sqrt{VT} \ll V, \\
{\displaystyle 1-e^{-\frac{\epsilon-V}{T} }  };   & \epsilon\gg V\gg T.
\end{array} \right.
                                                          \label{nug}
\end{equation}
The  exponential suppression of the tunneling DOS
is a direct manifestation of the Coulomb blockade. In the
extreme case of zero temperature  one has
\begin{equation}
\nu(\epsilon) \stackrel{T\rightarrow 0}{\longrightarrow}
\nu^{[0]}\theta \left( |\epsilon|-\frac{V}{2} \right).
                                                          \label{nui}
\end{equation}

To discuss the quasiparticle lifetime due to global charge fluctuations,
we need an explicit form of the electron Green
function. To this end we substitute the  spectral function of
non--interacting electrons,
$A^{[0]}_{\alpha}(\epsilon,\overline\mu)=
2\pi\delta(\epsilon-\epsilon_{\alpha}+\overline\mu)$, into Eq.\ (\ref{GM1}),
perform integrations and do analytical continuation
($\epsilon_n\rightarrow\epsilon+i\delta$). The result is
\begin{equation}
{\cal G}^{R}_{\alpha}(\epsilon,\mu)=-i \sqrt{\frac{\pi}{2VT}} \left[
f(\epsilon_{\alpha}-\overline\mu)
w \left(  \frac{x+V/2}{\sqrt{2VT}} \right)+
[1-f(\epsilon_{\alpha}-\overline\mu)]
w \left(  \frac{x-V/2}{\sqrt{2VT}} \right) \right],
                                                          \label{GRT}
\end{equation}
where $f(\epsilon)$ is the Fermi function,
$x=\epsilon-\epsilon_{\alpha}+\overline\mu+i\delta$ and $w(z)=
\exp\{-z^2\}\mbox{erfc}(-iz)$ is the error function \cite{Abramowitz}.
We note that the $w(z)$ does not have poles at finite energies. We
nevertheless may consider the width of the peak of the imaginary part of
${\cal G}^{R}$ as a measure of the quasiparticle life--time. As a result
\begin{equation}
\frac{1}{\tau_{qp}}\propto \sqrt{VT}.
                                                          \label{qp}
\end{equation}
Both the DOS and $\tau_{qp}$ are measurable in tunneling experiment
\footnote{Note that the AA result for the
dephasing time \cite{Altshuler85} scales as
$\frac{1}{\tau_{\phi}}\propto T^{2/(4-d)}D^{-d/(4-d)} $. For
$d=0$  it resembles our Eq.\ (\ref{qp}).}.

Finally we discuss the nature of $V\equiv V({\bf q}=0)$. Consider the
equivalent circuit shown in Fig. \ref{fg:2}. Its equilibrium noise spectrum
($T=0$) is given by the fluctuation--dissipation theorem
\begin{equation}
\langle U_{tot} U_{tot} \rangle = i\omega
\left( \frac{1}{i\omega C} + Z(\omega) \right).
                                                          \label{noise}
\end{equation}
The voltage fluctuations on the system (capacitor) are
$U_C=U_{tot}/(1+i\omega C Z)$, so the corresponding correlation function
is
\begin{equation}
K(\omega)\equiv e^2\langle U_C U_C \rangle =
\frac{e^2}{C}\, \frac{1}{1+ i\omega C Z(\omega)}.
                                                          \label{cf}
\end{equation}
We may now add to our original Hamiltonian, Eq. (\ref{H}), a Gaussian noise
term
$e U_C(t)\sum_{\alpha}a^{+}_{\alpha}a_{\alpha}$
(the correlator of the noise is given by Eq. (\ref{cf})).
Integrating over the environment noise and following
through the above derivation of ${\cal G}_{\alpha}$, we shall  have to
replace $V\rightarrow V-K(\omega_m)$ in Eq. (\ref{G1}). It is natural to
identify $V=e^2/C$, where $C$ is the total capacitance of the $d=0$ system.
In this case our noise--modified Coulomb boson (after Gaussian integration
over $\sigma_m$ in Eq. (\ref{G1})) reproduces the result of Refs.
\cite{Devoret91} and \cite{Glazman91} for fluctuations in ultrasmall
capacitance junctions.

We are indebted to A. Altland  and A. M. Finkelstein for the useful
suggestions. This research was supported by the German--Israel
Foundation (GIF) and the U.S.--Israel Binational Science Foundation (BSF).

{\em Note added after this work has been completed }: in a recent preprint by
Levitov and Shytov (SISSA N 9501130), these authors present an effective
semiclassical action through which they study tunneling into a $d=2$ system.
Their ${\bf q}=0$ limit agrees with our exact result.

\begin{figure}
\caption{ \label{fg:1}
The single particle Green function in the presence of the Coulomb
boson, ${\cal D}$.}
\end{figure}

\begin{figure}
\caption{\label{fg:2}
An equivalent circuit, consisting of the system's capacitance $C$ and
a noisy environment. }
\end{figure}

\end{document}